\documentclass{article}
\usepackage{tikz}
\usepackage{hyperref}
\usepackage{spconf,amsmath,amssymb,graphicx,booktabs}
\usepackage{multirow}
\usepackage{url}
\usepackage{mathtools}

\DeclareMathOperator{\diag}{diag}

\usepackage{amsmath,amssymb,bm}
\newcommand{\nrm}[1]{\lVert #1 \rVert}

\title{EquiSELD: Efficient training of equivariant sound event localization and detection networks }

\name{Goksenin Yuksel\textsuperscript{1}, Marcel van Gerven\textsuperscript{1}, Kiki van der Heijden\textsuperscript{1,2}}
\address{\textsuperscript{1}Dept. of Machine Learning and Neural Computing, Donders Institute,\\
Radboud University, Nijmegen, the Netherlands\\
\textsuperscript{2}Mortimer B. Zuckerman Institute, Columbia University, New York, USA}

\begin{document}
\ninept
\maketitle
\begin{abstract}
First-order Ambisonics (FOA) signals exhibit exact $\mathrm{O}(3)$ symmetry: The rotation or reflection of the FOA signal modifies the direction of arrival of the sound sources, while preserving the sound sources themselves. Prior attempts to utilize this spatial symmetry of FOA to improve the efficiency and robustness of sound event detection and localization (SELD) systems either learned only an approximation of the symmetry through rotation-based augmentation or relied on computationally expensive methods to integrate equivariance. Furthermore, prior work focused exclusively on $\mathrm{SO}(3)$ equivariance , leaving the potential of incorporating $\mathrm{O}(3)$ equivariance for SELD tasks unclear. To address these limitations, we developed \textsc{EquiSELD}. This equivariant attention network processes first-order Ambisonics as paired streams of $\mathrm{O}(3)$-invariant scalars and equivariant intensity vectors, producing an invariant activity magnitude and an equivariant DOA with a Multi-ACCDOA readout. To compare the impact of $\mathrm{O}(3)$ versus $\mathrm{SO}(3)$-equivariance, we designed a matched $\mathrm{SO}(3)$-only variant. \textsc{EquiSELD} outperforms prior equivariant networks on both simulated scenes with measured RIRs and recordings of real-world sound scenes at a fraction of the training cost. \textsc{EquiSELD} additionally surpasses the performance of non-equivariant SELD networks of a similar size on the simulated real-world sound scenes and achieves competitive performance on the real-world sound scenes.
\end{abstract}

\section{Introduction}

Sound Event Localization and Detection (SELD) provides a joint spatio-temporal characterization of acoustic scenes, supporting applications ranging from computational auditory scene analysis to immersive audio and smart environments~\cite{seld_1, seld_2}. First-order Ambisonics (FOA) has become the de facto standard input for SELD systems~\cite{PSELDNets, starss23, Ambisonics}. The mathematical definition of the FOA signal ensures that the sound scene's directional components transform equivariantly under any rotation or reflection of the scene, while its omnidirectional component (W) remains invariant. This spatial equivariance is a natural candidate to improve the efficiency and robustness of SELD systems~\cite{cgnets}.

Yet, most SELD systems process FOA with standard CNNs or Transformers that do not encode spatial equivariance directly, instead relying on rotational data augmentation to inject the spatial symmetry statistically~\cite{data_aug, STARSS23_SOTA, PSELDNets}. However, this approach consumes data and computational resources to learn a constraint that could be hard-coded and, even though such rotational transformations are physically exact in the Ambisonic domain, the trained network only approximates the desired symmetry.

Group-equivariant networks address this limitation by incorporating equivariance into the network directly. Sato et al.~\cite{cgnets} employed Clebsch–Gordan product networks~\cite{clebsch-gordan} to integrate $\mathrm{SO}(3)$ rotation equivariance into a unified rotation, scale, and time-translation framework. However, their architecture does not take into account reflection equivariance, is computationally expensive, and cannot represent overlapping instances of the same sound class (common in real-world acoustic scenes, ~\cite{starss23, TAU2021}). Outside the audio domain, equivariant attention networks have gained traction for processing sets of geometric features, combining the expressivity of transformers with exact symmetry guarantees~\cite{liao2023equiformer, se3} and building on the
permutation-equivariant set attention of~\cite{set_transformer}. A complementary line of work achieves equivariance through typed scalar--vector streams, in which invariant scalars gate equivariant vector channels via their norms~\cite{vector-neurons, vector-perceptron}.

Inspired by this direction, we propose, to the best of our knowledge, the first SELD network that is exactly equivariant to the full orthogonal group $\mathrm{O}(3)$. \textsc{EquiSELD}\footnote{Code\&Models:  \url{https://github.com/labhamlet/equiSELD}} is an equivariant attention network, processing each time--mel token as a pair of typed streams. Each token comprises $\mathrm{O}(3)$-invariant scalars and equivariant first-order intensity vectors. By computing attention weights and all nonlinearities solely from invariant scalars, and mixing vector channels via bias-free linear maps, we achieve exact equivariance. Our network accomodates overlapping sound sources of the same type, outputting Multi-ACCDOA~\cite{multi-accdoa} activity as an invariant magnitude vector and DOA as an equivariant unit vector. We furthermore isolate $\mathrm{O}(3)$'s contribution with a matched $\mathrm{SO}(3)$-only variant. 

The results demonstrate that \textsc{EquiSELD} produces exactly equivariant predictions, surpasses the performance of an existing $\mathrm{SO}(3)$-equivariant network~\cite{cgnets} with $2.9\times$ fewer parameters and $19\times$ less training time, and obtains a lower SELD score than SELDNets trained with and without exact $\mathrm{O}(3)$ data augmentation.

\section{Method}
\label{sec:method}



\begin{figure*}
    \centering
    \includegraphics[width=0.9\linewidth]{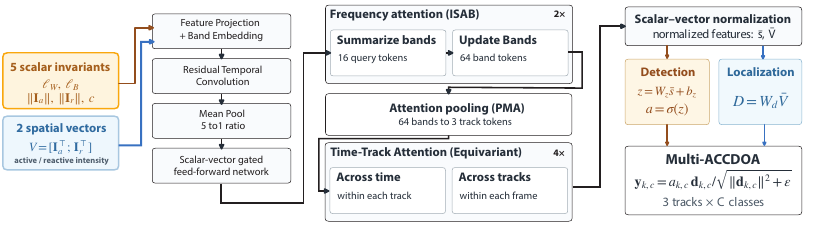}
    \caption{EquiSELD architecture. Five invariant scalar features (orange) and two equivariant intensity vectors (blue) are processed with frequency attention, attention pooling, 
    and attention across time and tracks. The network produces multi-ACCDOA outputs for $C$ sound classes for $K$ tracks of which direction estimates are equivariant. ISAB = induced set attention block; PMA = pooling by multihead attention~\cite{set_transformer}.}
    \label{fig:architecture}
\end{figure*}

\subsection{Invariant and Equivariant FOA Features}
\label{subsec:features}

First-order Ambisonics (FOA) comprises an omnidirectional channel $W$ and three directional channels $\mathbf B=[X,Y,Z]^{\top}$. Under rotations or reflections $Q\in O(3)$, $W$ is unchanged and $\mathbf B\mapsto Q\mathbf B$.

For each time--frequency bin of an STFT (1024-sample window, 20\,ms hop), we compute the complex intensity $\mathbf i=W^{*}\mathbf B$, its active and reactive components $\mathbf i_a=\Re\{\mathbf i\}$ and $\mathbf i_r=\Im\{\mathbf i\}$, and channel powers $E_W=|W|^2$ and $E_B=\nrm{\mathbf B}^2$. 

Features are pooled to 64 mel bands, with $\langle\cdot\rangle_m$ denoting the weighted frequency sum under the $m$-th mel filter. The normalized intensities are
\begin{equation}
\mathbf I_{p,m}=
\frac{\langle\mathbf i_p\rangle_m}{\bar E_m+\epsilon},
\qquad p\in\{a,r\},
\end{equation}
where $\bar E_m=\bigl\langle\tfrac12(E_W+\tfrac13E_B)\bigr\rangle_m$ is the mean band energy and $\epsilon=10^{-8}$ is used throughout for numerical stability.

Each time--mel bin contains two equivariant vectors, stacked as rows, and five invariant scalars:
\begin{equation}
\begin{aligned}
\mathbf V_m &=
\begin{bmatrix}
\mathbf I_{a,m}^{\top}\\
\mathbf I_{r,m}^{\top}
\end{bmatrix},\\
\mathbf s_m &=
\bigl[\ell_W,\ell_B,\nrm{\mathbf I_{a,m}},
\nrm{\mathbf I_{r,m}},c_m\bigr]^{\top}.
\end{aligned}
\end{equation}
Here $\ell_X=\log(\langle E_X\rangle_m+\epsilon)$, $X\in\{W,B\}$, are log band energies, and $c_m$ is the cosine similarity between $\mathbf I_{a,m}$ and $\mathbf I_{r,m}$

\subsection{Equivariant Token Blocks}
\label{subsec:ops}

A token $(\mathbf s,\mathbf V)$ pairs $d_s$ scalar channels with $d_v$ vector channels, stored as rows $\mathbf v_j^{\top}$ of $\mathbf V$. The group action is $Q\cdot(\mathbf s,\mathbf V)
=(\mathbf s,\mathbf VQ^{\top})$. A block $F$ is equivariant if
$F(Q\cdot\mathcal T)=Q\cdot F(\mathcal T)$
for every $Q\in O(3)$ and every token collection $\mathcal T$ .

Input tokens ($d_s=5$, $d_v=2$) are mapped to higher dimension by a scalar linear layer and a vector map~\eqref{eq:veclinear}. Throughout, $\mathbf W_{(\cdot)}$ and $\mathbf U$ are learnable real matrices; $\sigma(\cdot)$, $\diag(\cdot)$, $[\,\cdot\,;\,\cdot\,]$, and $\nrm{\cdot}_F$ denote the logistic sigmoid, a diagonal matrix, vertical concatenation, and the Frobenius norm.

\noindent\textbf{Vector linear maps.}
Vector channels are mixed without bias:
\begin{equation}
\mathcal L_v(\mathbf V)=\mathbf W_v\mathbf V.
\label{eq:veclinear}
\end{equation}
The same weights act on all three spatial coordinates, giving
$\mathcal L_v(\mathbf VQ^{\top})
=\mathcal L_v(\mathbf V)Q^{\top}$.

\noindent\textbf{Channel norms.}
The per-channel norms
\begin{equation}
\bm\nu(\mathbf V)=
\bigl[
\sqrt{\nrm{\mathbf v_1}^2+\epsilon},
\dots,
\sqrt{\nrm{\mathbf v_{d_v}}^2+\epsilon}
\bigr]^{\top}.
\end{equation}

satisfy $\bm\nu(\mathbf VQ^{\top})=\bm\nu(\mathbf V)$. Within the token blocks, vector information enters the scalar stream only through these invariant norms.

\noindent\textbf{Normalization.}
Scalars use LayerNorm~\cite{layer_norm}, denoted $\operatorname{LN}$. Vectors are normalized by their RMS channel norm~\cite{vector-perceptron} and rescaled by learnable channel gains $\bm\gamma$:
\begin{equation}
\mathcal N_v(\mathbf V)=
\frac{\diag(\bm\gamma)\mathbf V}
{\sqrt{d_v^{-1}\nrm{\mathbf V}_F^2+\epsilon}}.
\label{eq:vnorm}
\end{equation}
Since $\nrm{\mathbf VQ^{\top}}_F=\nrm{\mathbf V}_F$, the denominator is invariant and
$\mathcal N_v(\mathbf VQ^{\top})
=\mathcal N_v(\mathbf V)Q^{\top}$.

\noindent\textbf{Gated feed-forward block.}
An MLP produces scalar updates and one gate per vector channel:

\begin{equation}
\begin{aligned}
\widetilde{\mathbf V}
&=\mathcal L_{v,1}\bigl(\mathcal N_v(\mathbf V)\bigr),\\
\mathbf h
&=\operatorname{GELU}\bigl(
\mathbf W_1[\operatorname{LN}(\mathbf s);
\bm\nu(\widetilde{\mathbf V})]+\mathbf b_1\bigr),\\
[\Delta\mathbf s;\,\mathbf g]
&=\mathbf W_2\mathbf h+\mathbf b_2,\\
\mathbf s'&=\mathbf s+\Delta\mathbf s,\\
\mathbf V'&=\mathbf V+
\mathcal L_{v,2}\bigl(
\diag(\sigma(\mathbf g))\widetilde{\mathbf V}\bigr).
\end{aligned}
\end{equation}

where $\mathcal L_{v,1},\mathcal L_{v,2}$ follow~\eqref{eq:veclinear}, and $ \mathbf b_1,\mathbf b_2$ are learnable scalar biases. Both $\Delta\mathbf s$ and $\mathbf g$ are invariant. Applying each gate uniformly to a vector's three coordinates therefore preserves equivariance, as do the vector maps and residual addition.

\noindent\textbf{Equivariant attention.}
Following equivariant attention networks~\cite{se3}, we compute attention weights from invariant token descriptors:
\begin{equation}
\bm\phi_i=
\operatorname{LN}(\mathbf s_i)+
\mathbf U\,\bm\nu\bigl(\mathcal N_v(\mathbf V_i)\bigr).
\end{equation}
For one head with query/key width $d_h$, let
$\mathbf q_i=\mathbf W_q\bm\phi_i$ and
$\mathbf k_j=\mathbf W_k\bm\phi_j$. The attention weights are
\begin{equation}
A_{ij}=\operatorname{softmax}_j\!\left(
\frac{\mathbf q_i^{\top}\mathbf k_j}{\sqrt{d_h}}
\right).
\end{equation}
Here $i$ indexes query tokens and $j$ indexes key/value tokens. The same weights update both streams:
\begin{equation}
\begin{aligned}
\mathbf s_i'
&=\mathbf s_i+\mathbf W_o^s
\sum_j A_{ij}\mathbf W_v^s\bm\phi_j,\\
\mathbf V_i'
&=\mathbf V_i+\mathcal L_o^v\Bigl(
\sum_j A_{ij}\mathcal L_v^v
\bigl(\mathcal N_v(\mathbf V_j)\bigr)\Bigr).
\end{aligned}
\end{equation}
The matrices $\mathbf W_v^s,\mathbf W_o^s$ are scalar value and output projections, and $\mathcal L_v^v,\mathcal L_o^v$ are their vector counterparts~\eqref{eq:veclinear}.

Because $\bm\phi_i$ is invariant, queries, keys, and attention weights are unchanged under $Q$. For vector values $\mathbf Z_j = \mathcal L_v^v
\bigl(\mathcal N_v(\mathbf V_j)\bigr)$, aggregation satisfies the equivariance constraint. Channel concatenation, residual addition, and composition preserve this property, so all blocks above are equivariant to rotations and reflections.

\subsection{SELD Architecture}
\label{subsec:architecture}

\textsc{EquiSELD} adapts Set Transformer attention~\cite{set_transformer} to paired scalar--vector features, using $d_s=128$, $d_v=32$, and 2.25M parameters. Figure~\ref{fig:architecture} specifies layer order, token counts, and block repetitions. Temporal vector convolutions are bias-free and share kernels across Cartesian components. We add learnable mel-band positional embeddings to the scalar stream, shared across time. Frequency attention and pooling follow the induced set attention block (ISAB) and pooling by multihead attention (PMA), respectively. ISAB/PMA seeds have learnable scalar components only, with seed vectors fixed at zero. Time--track attention operates across time within each track and across tracks within each frame. Output heads combine invariant activities with normalized equivariant directions into multi-ACCDOA outputs~\cite{multi-accdoa} for $C$ classes and $K$ tracks per class.

\section{Experiments}
\label{sec:experiments}

\begin{table*}[!t]
  \centering
  \scriptsize
  \setlength{\tabcolsep}{3pt}
  \caption{SELD performance on STARSS23 and TAU2021, with training cost on
  TAU2021 (single A100). For STARSS23, ``Det.'' is the number of the 13
  classes the model detects and the parenthesised LE is the
  class-averaged localization error restricted to those classes. We show the jackknife estimates of the $\mathcal{E}_{\mathrm{S}}$ under the test-set recordings.}
  \label{tab:main}
  \begin{tabular}{lrr ccccc c ccccc}
    \toprule
    & & & \multicolumn{6}{c}{STARSS23} & \multicolumn{5}{c}{TAU2021} \\
    \cmidrule(lr){4-9}\cmidrule(lr){10-14}
    Model & Par. & GPU h$\downarrow$ & ER$\downarrow$ & F$\uparrow$ & LE (det.)$\downarrow$ & LR$\uparrow$ & Det.$\uparrow$ & $\mathcal{E}_{\mathrm{S}} \downarrow$
                 & ER$\downarrow$ & F$\uparrow$ & LE$\downarrow$ & LR$\uparrow$ & $\mathcal{E}_{\mathrm{S}}$$\downarrow$ \\
    \midrule
    SELDNet & 0.75M & 0.16
      & 0.74 & 13.7 & $55.7^\circ$ ($33.1^\circ$) & 26.5 & 11 & 0.66 {[0.61, 0.69]}
      & 0.68 & 27.5 & $35.8^\circ$ & 42.8 & 0.55 {[0.53, 0.56]} \\
    SELDNet & 2.25M & 0.18
      & 0.75 & 16.1 & $51.9^\circ$ ($28.6^\circ$) & 29.5 & 11 & 0.65 {[0.62, 0.68]}
      & 0.68 & 29.8 & $22.3^\circ$ & 46.6 & 0.51 {[0.50, 0.52]} \\
    SELDNet$^{+}$ & 2.25M & 0.83
      & \textbf{0.57} & \textbf{25.4} & $68.2^\circ$ ($18.5^{\circ}$) & 30.0 & \phantom{0}9 & 0.60 {[0.57, 0.63]}
      & 0.61 & 40.5 & $32.3^\circ$ & 49.4 & 0.47 {[0.46, 0.49]} \\
    CGNet-STS-MultiACCDOA & 6.53M & 14.5
      & 0.80 & 11.3 & $78.3^\circ$ ($33.1^\circ$) & 24.5 & \phantom{0}9 & 0.72 {[0.68, 0.74]}
      & 0.73 & 30.1 & $25.3^\circ$ & 52.0 & 0.51 {[0.50, 0.53]} \\
    CGNet-STS & 6.28M & 12.1
      & -- & -- & -- & -- & -- & --
      & 0.79 & 26.2 & $31.9^\circ$ & 55.1 & 0.54 {[0.52, 0.55]} \\
    \midrule
    EquiSELD& 2.25M & 0.77
      & 0.72 & 24.9 & $34.3^\circ$ ($22.1^\circ$) & 39.7 & \textbf{12} & \textbf{0.56} {[0.51, 0.59]}
      & 0.64 & 49.8 & 16.4$^\circ$ & \textbf{63.9} & 0.40 {[0.38, 0.41]} \\
    EquiSELD-SO(3) & 2.25M & 0.77
      & 0.69 & 21.4 & $\mathbf{31.7^\circ}$ ($19.4^\circ$) & 30.9 & 12 & 0.58 {[0.50, 0.61]}
      & \textbf{0.60} & \textbf{51.3} & $\mathbf{15.9^\circ}$ & \textbf{63.9} & \textbf{0.38} {[0.37, 0.40]} \\
    AttnSELD & 2.25M & 0.77
      & 0.97 & 13.6 & $52.6^\circ$ ($29.5^\circ$) & 30.6 & 11 & 0.70 {[0.68, 0.74]}
      & 0.71 & 42.6 & $17.9^\circ$ & 57.6 & 0.45 {[0.44, 0.47]} \\
    AttnSELD$^{+}$& 2.25M & 0.85
      & 0.82 & 25.1 & 33.6$^\circ$ ($21.5^\circ$) & \textbf{44.9} & \textbf{12} & 0.58 {[0.51, 0.61]}
      & 0.67 & 42.2 & $18.8^\circ$ & 56.7 & 0.45 {[0.43, 0.46]} \\
    \bottomrule
  \end{tabular}
\end{table*}
\subsection{Setup}
\label{sec:setup}

\textbf{Datasets.} We evaluate our method on two established SELD tasks: TAU2021~\cite{TAU2021} and STARSS23~\cite{starss23}. TAU2021 is a synthetically generated dataset based on real recorded room impulse responses (RIRs) featuring moving sound sources and interfering noise (sampling rate = 24 kHz). It comprises six folds, each containing approximately 100 minutes of audio. We use folds 1--4 for training, fold 5 for validation, and fold 6 for testing. STARSS23 is a recorded dataset of real-world listening scenes that presents a more challenging acoustic environment with up to six overlapping sound events (sampling rate = 24 kHz). All STARSS23 systems are trained on the real dev-train split only.

\noindent\textbf{Training and evaluation.} We adopted the Multi-ACCDOA output format with $K=3$ tracks for each of $C$ sound classes and trained models on 5 s segments using the ADPIT loss~\cite{multi-accdoa} and a detection threshold of $0.5$. We trained models for 100 epochs on TAU2021 and for 200 epochs for STARSS23. We used a longer training regime for STARSS23 to avoid underfitting.

\noindent\textbf{Evaluation metrics.} Following the conventional evaluation of SELD systems, we report the macro-averaged,
location-dependent error rate ($\mathrm{ER}_{20^\circ}$) and F-score
($\mathrm{F}_{20^\circ}$). Here, a prediction counts as a true positive
only if the DOA falls within $20^\circ$ of the target. We furthermore report the
class-dependent localization error ($\mathrm{LE}_{\mathrm{CD}}$) in
degrees and localization recall ($\mathrm{LR}_{\mathrm{CD}}$)~\cite{dcase_metrics}. We use the $\mathcal{E}_{\mathrm{S}}$ to quantify overall SELD performance:
\begin{equation}
    \mathcal{E}_{\mathrm{S}} = \frac{1}{4}
[\mathrm{ER}_{20^\circ} + (1 - \mathrm{F}_{20^\circ}) +
\mathrm{LE}_{\mathrm{CD}}/180^\circ + (1 - \mathrm{LR}_{\mathrm{CD}})].
\end{equation}

\subsection{Model space and training parameters}
\label{sec:models}

\noindent\textbf{EquiSELD.} We optimized \textsc{EquiSELD} with AdamW~\cite{adamw} ($\mathrm{lr} = 3\times10^{-4}$, $\beta = (0.9, 0.95)$, and weight decay $0.05$). We applied gradient clipping at $1.0$ and used a linear warmup over 2000 steps followed by cosine decay down to $1\%$ of the peak learning rate. 

\noindent\textbf{EquiSELD-SO(3).} To quantify the potential benefit of $O(3)$ equivariance of \textsc{EquiSELD} in comparison with $SO(3)$ equivariance, we introduce an $SO(3)$-equivariant
version of the model by mixing a third, axial-vector channel --- the cross product
$\mathbf{v}_a \times \mathbf{v}_r$ --- into the vector stream. As the product of two
polar vectors, this feature transforms as $\mathbf{v} \mapsto \det(Q)\,Q\mathbf{v}$. This is identical to a polar vector under proper rotations, but sign-flipped under reflection. It therefore reduces the guaranteed symmetry from $\mathrm{O}(3)$ to $\mathrm{SO}(3)$. The variant is otherwise
identical to EquiSELD in architecture, parameter count, and training recipe.

\noindent\textbf{AttnSELD.} To compare equivariant and non-equivariant parameterizations at
matched model size, we construct \textsc{AttnSELD}. This architecture is similar to \textsc{EquiSELD} but releases the constraints that guarantee equivariance through four modifications: (i) vector
channel mixings gain learned $3\times3$ spatial mixing matrices
and per-channel biases; (ii) attention logits and the feed-forward
block read raw three-dimensional vector components rather than
their norms; (iii) vector normalization permits per-component
affine transformations; and (iv) inducing points and the pooling
head carry learnable, non-zero vector parts. Since exposing raw
components increases the input dimensions of the scalar
projections, we adjust the vector width to $d_v=12$ ($d_v=32$ in \textsc{EquiSELD}) resulting in nearly 2.25 M parameters
as in \textsc{EquiSELD}. 

\noindent\textbf{AttnSELD+.} To assess to what extent the non-equivariant version of \textsc{EquiSELD} can learn equivariance through
$\mathrm{O}(3)$ augmentation, we introduce \textsc{AttnSELD+}. Our augmentation implementation follows the channel-first ~\cite{data_aug}. We sample an element from $\mathrm{O}(3)$ for each instance in the batch and apply on-the-fly augmentations. It is continuous over the O(3) elements.

\noindent\textbf{SELDNet.} We adopt the SELDNet variant that served as the official DCASE 2023 Task 3 baseline~\cite{starss23}. The CRNN of~\cite{seldnet}, extended with two multi-head self-attention layers and a Multi-ACCDOA output~\cite{multi-accdoa}. This model first extracts a concatenated mel-spectrogram and normalized active intensity vector (IV) using a CNN front-end, then processes the resulting frame-wise features with a GRU followed by multi-head self-attention. We train two variants: the conventional recipe with 0.75 M parameters, and a larger parameter-matched version with 2.25 M parameters (obtained by increasing the GRU hidden dimension from 128 to 240). All SELDNet variants follow the published training recipe using the Adam optimizer~\cite{adam} and a learning rate of $10^{-3}$. We also introduce \textbf{SELDNet+}, which follows the same data augmentation schema as the \textsc{AttnSELD+}.

\noindent\textbf{CGNet-STS and CGNet-STS-MultiACCDOA.} CGNet-STS is an existing SELD network that incorporates equivariance to $\mathrm{SO}(3)$ rotations, time translations, and scale changes through Clebsch–Gordan product networks~\cite{clebsch-gordan}. We re-implemented the architecture from the original paper and verified functional equivalence against the authors' reference code\footnote{\url{https://github.com/nttrd-mdlab/group-equiv-seld}}. However, the original implementation outputs separate DOA and SED predictions, which prevents the model from handling overlapping sound events of the same class. We therefore adapted CGNet-STS to the Multi-ACCDOA framework by widening its final layer to emit $K \times C$ first-order direction vectors (where $K$ is the number of tracks and $C$ is the number of classes) and widening the GRU readout to produce matching logits $z_{k,c}$. The final Multi-ACCDOA vector is then formed as $\mathbf{a}_{k,c} = \sigma(z_{k,c})\,\mathbf{u}_{k,c} / \lVert \mathbf{u}_{k,c} \rVert$. As we found that the Multi-ACCDOA version performs better on the synthetic scenes than the original implementation (Section~\ref{sec:results}), we continued with this version for real-world STARSS23 dataset.

\section{Results}
\label{sec:results}

\subsection{SELD Results}
\label{subsec:seld_results}

On the simulated scenes with measured RIRs (TAU2021), EquiSELD achieves a summary SELD score 
of $\mathcal{E}_{\mathrm{S}} = 0.40$, outperforming all versions of SELDNet (lowest $\mathcal{E}_{\mathrm{S}} = 0.47$) and CGNet-STS (lowest $\mathcal{E}_{\mathrm{S}} = 0.51$, Table ~\ref{tab:main}). Crucially, \textsc{EquiSELD} has
$2.9\times$ fewer parameters and uses approximately $19\times$ less GPU hours than CGNet-STS to achieve this performance, showcasing the effectiveness of equivariant parametrization via the equivariant transformers framework. Finally, on the more challenging and limited real-world scenes (STARSS23), EquiSELD achieves a summary SELD score 
of $\mathcal{E}_{\mathrm{S}} = 0.56$, once more surpassing all versions of SELDNet (lowest $\mathcal{E}_{\mathrm{S}} = 0.60$) and CGNet-STS ($\mathcal{E}_{\mathrm{S}} = 0.72$).  

\noindent\textbf{Rotation vs.\ rotation and reflection equivariance.}
\textsc{EquiSELD-SO(3)} achieves comparable summary SELD scores as \textsc{EquiSELD} under $O(3)$ equivariance for both the synthetic and real-world scenes (Table ~\ref{tab:main}, see overlapping jackknife confidence intervals). These results indicate that the gain in effectiveness of $O(3)$ equivariance (rotation and reflection equivariance) is comparable to the gain of effectiveness of $SO(3)$ equivariance (rotation equivariance only). 

\noindent\textbf{EquiSELD without equivariance constraints.} \textsc{AttnSELD} performs substantially worse on the synthetic scenes (TAU2021, $\mathcal{E}_{\mathrm{S}} = 0.45$) as well as the real-world scenes (STARSS23, $\mathcal{E}_{\mathrm{S}} = 0.70$, Table ~\ref{tab:main}). These findings highlight the impact of the equivariance constraints on model performance. Training \textsc{AttnSELD+} with $O(3)$ augmentation improved performance on the real-world scenes at a small increase of the training cost, achieving a similar summary score as \textsc{EquiSELD} ($\mathcal{E}_{\mathrm{S}} = 0.58$, Table ~\ref{tab:main}).  However, performance of \textsc{AttnSELD+} did not improve on the synthetic scenes ($\mathcal{E}_{\mathrm{S}} = 0.45$, Table ~\ref{tab:main}).

\noindent\textbf{Undetected classes. }On STARSS23, $\mathrm{LE}_{\mathrm{CD}}$ assigns $180^\circ$
to undetected classes, coupling class coverage with localization
precision. SELDNet$^{+}$ and CGNet-STS each detect nine of 13
classes, yielding overall errors of $68.2^\circ$ and $78.3^\circ$,
respectively, versus $18.5^\circ$ and $33.1^\circ$ over detected
classes. These restricted averages involve different class
subsets and are therefore not directly comparable.

\subsection{Orientation Robustness}
\label{subsec:robustness}

We evaluate the spatial equivariance of each model on a sweep of 74 elements of $\mathrm{O}(3)$: 12
Fibonacci-sphere directions with 3 in-plane rolls each, plus the identity, paired with their parity partners $-Q$. On test-fold recordings comparing $f(Qx)$ against $Qf(x)$, EquiSELD achieves exact equivariance (maximum deviation for \textsc{EquiSELD} = $2.8\times10^{-7}$). Table~\ref{tab:o3_sweep} shows that SELDNet's scores range from 0.65 to 0.82 and AttnSELD's range from 0.70 to 0.86. Variants with $O(3)$ augmentations exhibit smaller residual variation as well, but EquiSELD attains this without augmentations and gives the best absolute performance.

\begin{table}[!htbp]
  \centering
  \scriptsize
  \caption{SELD score over O(3) elements on STARSS23.}
  \label{tab:o3_sweep}
  \begin{tabular}{lcccc}
    \toprule
    Model & Mean & Std & Min & Max \\
    \midrule
    SELDNet & 0.76 & $\pm$0.03 & 0.65 & 0.82 \\
    SELDNet$^{+}$ & 0.60 & $\pm$0.01 & 0.60 & 0.62 \\ 
    AttnSELD & 0.75 & $\pm$0.04 & 0.70 & 0.86 \\
    AttnSELD$^{+}$ & 0.58 & $\pm$0.01 & 0.57 & 0.59 \\
    EquiSELD (ours)   & \textbf{0.56} & $\pm$\textbf{0.00} & 0.56 & 0.56 \\
    \bottomrule
  \end{tabular}
\end{table}



\begin{figure}[!htbp]
    \centering
\includegraphics[width=0.8\linewidth]{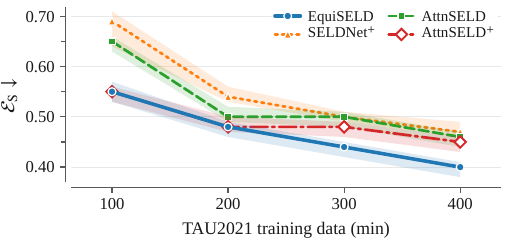}
    \caption{Data scaling. Shaded areas are
95\% confidence intervals.}
    \label{fig:scaling}
\end{figure}

\subsection{Data Scaling}
\label{ssec:scaling}

To compare how the effectiveness of inherent network equivariance scales with data quantity, we depict performance on TAU2021 as the training set increases from one to four folds ($100$--$400$ min, Figure~\ref{fig:scaling}). \textsc{EquiSELD} improves monotonically from $\mathcal{E}_{\mathrm{S}} = 0.55$ to $\mathcal{E}_{\mathrm{S}} = 0.40$, while SELDNet+, \textsc{AttnSELD} and \textsc{AttnSELD+} show less improvement. This graph shows that for simulated scenes with measured RIRs, \textsc{EquiSELD} outperforms \textsc{AttnSELD+}, especially at larger training-set sizes.



\section{Conclusion}
We proposed \textsc{EquiSELD}, an \(O(3)\)-equivariant attention network that encodes the rotation and reflection symmetry. The inherent equivariance of this low-parameter network requires no $O(3)$ augmentation to build rotation and reflection robustness. Experimental results show that \textsc{EquiSELD} outperforms a prior equivariant SELD network and the SELD baseline system on both synthetic and real-world scenes. Moreover, \textsc{EquiSELD} has \(2.9\times\) fewer parameters and requires \(19\times\) less
training time than the preceding equivariant network. Finally, we show that the benefit of inherent equivariance scales with data size, consistent with findings
on scaling equivariant transformers more broadly~\cite{scaling-equiv}. Future work will explore softening this hard
constraint --- through soft-equivariance penalties --- to exploit the elevation priors that strict
\(O(3)\) and \(SO(3)\) symmetry cannot, and extending the network to higher-order Ambisonics.

\section{Acknowledgments}
This work used the Dutch national e-infrastructure with the support of the SURF Cooperative using grant no. EINF-14624.

\bibliographystyle{IEEEbib}
\bibliography{strings,refs}

\end{document}